\documentclass[preprint,review,showpacs,preprintnumbers,amsmath,amssymb,12pt]{elsarticle}
\usepackage{txfonts}
\usepackage{amssymb}
\usepackage{tipa}
\usepackage{graphicx}
\usepackage{amssymb}

\journal{Optics Communications}

\begin{document}
\begin{frontmatter}

\title{Joint remote preparation of a four-dimensional quantum state}

\author{You-Bang Zhan\corref{cor1}}
\cortext[cor1]{\emph{E-mail address}: ybzhan@hytc.edu.cn}

\address{School of Physics and Electronic Electrical Engineering, Huaiyin Normal University, Huaian 223300,
PR China}

\begin{abstract}
We propose various protocols for joint remotely prepare a
four-dimensional quantum state by using two- and three-particle
four-dimensional entangled state as the quantum channel. The single-
and two-particle generalized projective measurement and the
appropriate unitary operation are needed  in our protocols. It is
shown that the receiver can reconstruct the unknown original state
only if two senders collaborate with each other. \bigskip \bigskip
\end{abstract}


\end{frontmatter}
\newpage

\section{Introduction}
Quantum entanglement plays a more and more critical role in quantum
information theory. Quantum teleportation, proposed by Bennett
\emph{et al}[1], is the process that transmits an unknown quantum
state from a sender to spatially distant receiver using the
entanglement channel with the help of some classical information. In
the last decade, Lo[2], Pati[3], and Bennett \emph{et al}[4]
presented a new quantum communication protocol that uses classical
communication and a previously shared entangled resource to remotely
prepare a quantum state. This communication protocol is called
remote state preparation(RSP). RSP is another important protocol
taking advantage of entanglement, in which the sender Alice performs
a measurement on her share of the entangled resource in a basis
chosen in accordance with the state she wishes to help the receiver
Bob in his laboratory to prepare. In RSP, Alice is assumed to know
fully the transmitted state to be prepared by Bob, so RSP is called
the teleportation of a known state. Compared with teleportation. RSP
requires less classical communication cost than teleportation[3]. In
recent years, RSP has attracted much attention, various
protocols[5-12] for generalization of RSP have been presented using
various kinds of methods, including low-entanglement RSP[5], optimal
RSP[6], generalized RSP[7], oblivious RSP[8], continuous variable
RSP[9,10], etc. Several RSP protocols in higher dimensional Hilbert
space have been proposed[13-15]. Meanwhile, some RSP protocols have
already been implemented experimentally[16-20].

All the above protocols assume the case that only one sender knows
the original state. However, if two-party or multiparty share an
original quantum state, and they want to remotely prepare it in the
laboratory of receiver, how can they do it ? To answer this problem,
recently, a novel aspect of RSP, called as the joint remote state
preparation(JRSP), has been proposed[21-25]. In these protocols of
the JRSP[21-25], two senders(or N senders) know partly of the
original state they want to remotely prepare, respectively. If and
only if all the senders agree to collaborate, the receiver can
reconstruct the original quantum state. Nevertheless, in
Refs.[21-25] only the single- or multi-qubit state was considered.
Though various protocols of RSP of high-dimensional quantum state
have been proposed in recent years[13-15], but no scheme has yet
been reported for the JRSP of higher-dimensional quantum state.

During the last few years, the high-dimensional system in quantum
information processing(QIP) has attracted much attention.
High-dimensional systems have properties which are different from
qubit counterparts which could be useful for QIP. For instance, we
can note that high-dimensional systems can be more entangled than
qubits[26-28] and can share a larger fraction of their
entanglement[29]. These properties, as well as the larger dimension
alone could aid many QIP tasks, including quantum key
distribution[30-33], quantum teleportation[1,34], quantum bit
commitment[35,36], quantum computing[37-41], quantum dense
coding[42], quantum secure communication[43], quantum secret
sharing[44,45], and quantum state remotely preparation[13-15]. In
this paper, we propose a set of protocols for two senders to
remotely prepare the single- and two-particle four-dimensional(FD)
quantum state by using various types of quantum channel. In our
protocols, the single- and two-particle FD projective measurement
and appropriate unitary operation are needed. This paper is
organized as follows. In section 2, a protocol for joint remotely
prepare an unknown single-particle FD quantum state by using a
tripartite FD entangled state as quantum channel is presented. In
section 3, we propose two protocols of joint remote preparation of
an unknown bipartite FD entangled state via two tripartite and three
bipartite FD entangled states as quantum channel, respectively.
Conclusions are given in section 4.

\section{Joint remote preparation of a single-particle FD quantum state}
Suppose that Alice$_{1}$, and Alice$_{2}$ share the original state
$|\varphi\rangle$, and they wish to help Bob remotely prepare a FD
quantum state
\begin{equation}\label{eq1}
|\varphi\rangle=\alpha|0\rangle+\beta|1\rangle+\gamma|2\rangle+\delta|3\rangle,
\end{equation}
like with Ref.[13-15], here we only consider that $\alpha$, $\beta$,
$\gamma$ and $\delta$ are real and
$\alpha^2+\beta^2+\gamma^2+\delta^2=1$. Suppose that Alice$_{1}$ and
Alice$_{2}$ know the original state $|\varphi\rangle$ partly, that
is, Alice$_{1}$ knows $\alpha_{1}$, $\beta_{1}$, $\gamma_{1}$ and
$\delta_{1}$, Alice$_{2}$ knows $\alpha_{2}$, $\beta_{2}$,
$\gamma_{2}$ and $\delta_{2}$, where $\alpha_{1}\alpha_{2}=\alpha$,
$\beta_{1}\beta_{2}=\beta$, $\gamma_{1}\gamma_{2}=\gamma$ and
$\delta_{1}\delta_{2}=\delta$. We also assume that the quantum
channel shared by Alice$_{1}$, Alice$_{2}$ and Bob is the tripartite
FD entangled state
\begin{equation}\label{eq2}
|\phi\rangle_{123}=\frac{1}{2}\sum_{j=0}^{3}|jjj\rangle_{123}.
\end{equation}
Assume particle 1 belongs to Alice$_{1}$, particle 2 to Alice$_{2}$,
and particle 3 to Bob, respectively. In order to help Bob remotely
prepare the original state, what Alice$_{1}$ and Alice$_{2}$ need to
do is to perform  single-particle FD projective measurements on
their own particles 1 and 2 respectively. The measurement basis
chosen by Alice$_{1}$ and Alice$_{2}$ are the set of mutually
orthogonal basis vectors (MOBVs) $\{|\psi_{0}^{(k)}\rangle,
\{|\psi_{1}^{(k)}\rangle, \{|\psi_{2}^{(k)}\rangle,
\{|\psi_{3}^{(k)}\rangle\}$ which is given by
\begin{eqnarray}\label{eq3}
|\psi_{0}^{(k)}\rangle=\alpha_{k}|0\rangle+\beta_{k}|1\rangle+\gamma_{k}|2\rangle+\delta_{k}|3\rangle,\nonumber\\
|\psi_{1}^{(k)}\rangle=\beta_{k}|0\rangle-\alpha_{k}|1\rangle+\delta_{k}|2\rangle-\gamma_{k}|3\rangle,\nonumber\\
|\psi_{2}^{(k)}\rangle=-\gamma_{k}|0\rangle+\delta_{k}|1\rangle+\alpha_{k}|2\rangle-\beta_{k}|3\rangle,\nonumber\\
|\psi_{3}^{(k)}\rangle=-\delta_{k}|0\rangle-\gamma_{k}|1\rangle+\beta_{k}|2\rangle+\alpha_{k}|3\rangle,
\end{eqnarray}
where $k=1$ and 2, and $\{|\psi_{j}^{(1)}\rangle\}$ $(j=0,1,2,3)$ is
a set of MOBVs chosen by Alice$_{1}$, and
$\{|\psi_{j}^{(2)}\rangle\}$ by Alice$_{2}$. Since $\alpha_{1}$,
$\beta_{1}$, $\gamma_{1}$ and $\delta_{1}$ ($\alpha_{2}$,
$\beta_{2}$, $\gamma_{2}$  and  $\delta_{2}$) that are necessary for
determining the basis $\{|\psi_{j}^{(1)}\rangle\}$
$(\{|\psi_{j}^{(2)}\rangle\})$ are known only to Alice$_{1}$
(Alice$_{2}$), so Alice$_{1}$ and Alice$_{2}$ are always able to
make the measurements independently of each other.

By Eq.(3), the state (2) can be described as
\begin{eqnarray}\label{eq4}
|\phi\rangle_{123}=\frac{1}{2}[|\psi_{0}^{(1)}\rangle_{1}|\psi_{0}^{(2)}\rangle_{2}(\alpha|0\rangle+\beta|1\rangle+\gamma|2\rangle+\delta|3\rangle)_{3}\qquad\qquad\qquad\qquad\nonumber\\
+|\psi_{1}^{(1)}\rangle_{1}|\psi_{1}^{(2)}\rangle_{2}(\beta|0\rangle+\alpha|1\rangle+\delta|2\rangle+\gamma|3\rangle)_{3}\qquad\qquad\nonumber\qquad\qquad\\
+|\psi_{2}^{(1)}\rangle_{1}|\psi_{2}^{(2)}\rangle_{2}(\gamma|0\rangle+\delta|1\rangle+\alpha|2\rangle+\beta|3\rangle)_{3}\qquad\qquad\nonumber\qquad\qquad\\
+|\psi_{3}^{(1)}\rangle_{1}|\psi_{3}^{(2)}\rangle_{2}(\delta|0\rangle+\gamma|1\rangle+\beta|2\rangle+\alpha|3\rangle)_{3}\qquad\qquad\nonumber\qquad\qquad\\
+|\psi_{0}^{(1)}\rangle_{1}|\psi_{1}^{(2)}\rangle_{2}(\alpha_{1}\beta_{1}|0\rangle-\beta_{1}\alpha_{2}|1\rangle+\gamma_{1}\delta_{2}|2\rangle-\delta_{1}\gamma_{2}|3\rangle)_{3}+\cdots
\ ],
\end{eqnarray}
where "$\cdots$" represents 11 terms with $l \neq m$ in
$|\psi_{l}^{(1)}\rangle_{1}|\psi_{m}^{(2)}\rangle_{2}$ $(l, m
=0,1,2,3)$. Clearly, in Eq.(4) only the four first terms can cause
success, but all the 12 remaining terms with $l\neq m$ lead to
failure. Now let Alice$_{1}$ and Alice$_{2}$ perform single-particle
FD projective measurements on their own particles 1 and 2
respectively, and then they inform Bob of their results by the
classical channels. According to the measurement results of
Alice$_{1}$ and Alice$_{2}$, the receiver Bob can reconstruct the
original state at his side. Without loss of generality, assume
Alice$_{1}$'s measurement result is $|\psi_{1}^{(1)}\rangle_{1}$ and
Alice$_{2}$'s result is $|\psi_{1}^{(2)}\rangle_{2}$, the particle 3
will collapse into the state
$\frac{1}{2}(\beta|0\rangle+\alpha|1\rangle+\delta|2\rangle+\gamma|3\rangle)_{3}$.
Bob needs to perform a local unitary operation $U_{1}$ on particle
4, the state of particle 3 will evolve
$\frac{1}{3}(\alpha|0\rangle+\beta|1\rangle+\gamma|2\rangle+\delta|3\rangle)_{3}$,
which is exactly the original state $|\varphi\rangle$. Here the
unitary operation $U_{1}$ is one of the unitary operations $U_{i}
(i=0,1,2,3)$
\begin{eqnarray}\label{eq5}
U_{0}=\left(
\begin{array}{cc}
 I & 0 \\
 0 & I
\end{array}
\right),\, U_{1}=\left(
\begin{array}{cc}
 \sigma_{x} & 0 \\
 0 & \sigma_{x}
\end{array}
\right),\,  U_{2}=\left(
\begin{array}{cc}
 0 & I \\
 I & 0
\end{array}
\right),\, U_{3}=\left(
\begin{array}{cc}
 0 & \sigma_{x} \\
\sigma_{x} & 0
\end{array}
\right),
\end{eqnarray}
where $I$ is $2\times2$ identity matrix and $\sigma_{x}$ is Pauli
matrix. If the measurement results of Alice$_{1}$ and Alice$_{2}$
are the other 3 cases of the four first terms in Eq.(4), the
relation between the results obtained by Alice$_{1}$ and Alice$_{2}$
and the unitary operations performed by Bob is shown in Table 1. The
required classical communication cost is 4 bits $(2\times\log_{2}4)$
in the protocol.
\begin{table}
\caption{Corresponding relation between the measurement results(MR)
of Alice$_{1}$ and Alice$_{2}$ and unitary operation $U_{i}$ by Bob}
\label{tab:1}
\begin{center}
\begin{tabular}{cc}
\hline
MR & \qquad\quad $U_{i}$\\
\hline \quad$|\psi_{0}^{(1)}\rangle_{1}|\psi_{0}^{(2)}\rangle_{2}$ & \quad\qquad $U_{0}$\\
\quad$|\psi_{1}^{(1)}\rangle_{1}|\psi_{1}^{(2)}\rangle_{2}$ &  \qquad\quad $U_{1}$\\
\quad$|\psi_{2}^{(1)}\rangle_{1}|\psi_{2}^{(2)}\rangle_{2}$ &  \qquad\quad $U_{2}$\\
\quad$|\psi_{3}^{(1)}\rangle_{1}|\psi_{3}^{(2)}\rangle_{2}$ &  \qquad\quad $U_{3}$\\
\hline
\end{tabular}
\end{center}
\end{table}

\section{Joint remote preparation of a bipartite FD entangled state}
We now consider the situation when the state of joint remote
preparation is a bipartite FD entangled state. In what follows we
present two protocols of JRSP using different quantum resources as
the quantum channel. The first protocol relies two tripartite FD
entangled states and the second protocol uses three bipartite FD
entangled states as the quantum channel, respectively.

\subsection{JRSP by using two tripartite FD entangled states as the quantum channel}
Suppose that Alice$_{1}$ and Alice$_{2}$ wish to help the receiver
Bob remotely prepare a bipartite FD entangled state
\begin{equation}\label{eq6}
|\psi\rangle=a|00\rangle+b|11\rangle+c|22\rangle+d|33\rangle,
\end{equation}
where $a$, $b$, $c$ and $d$ are real and $a^2+b^2+c^2+d^2=1$. Assume
that Alice$_{1}$ and Alice$_{2}$ know the original state
$|\psi\rangle$ partly, i.e., Alice$_{1}$ knows $a_{1}$, $b_{1}$,
$c_{1}$ and $d_{1}$, Alice$_{2}$ knows $a_{2}$, $b_{2}$, $c_{2}$ and
$d_{2}$, where $a_{1}a_{2}=a, b_{1}b_{2}=b, c_{1}c_{2}=c$ and
$d_{1}d_{2}=d$. We also suppose that the quantum channels shared by
Alice$_{1}$, Alice$_{2}$ and Bob are two tripartite FD entangled
states
\begin{eqnarray}\label{eq7}
|\phi\rangle_{123}=\frac{1}{2}\sum_{j=0}^{3}|jjj\rangle_{123},\nonumber\\
|\phi\rangle_{456}=\frac{1}{2}\sum_{j=0}^{3}|jjj\rangle_{456}.
\end{eqnarray}
Here, particles 1 and 4 belong to Alice$_{1}$, particles 2 and 5 to
Alice$_{2}$ and particle 3 and 6 to Bob, respectively. In order to
help Bob to remotely prepare the original state, Alice$_{1}$ and
Alice$_{2}$ should perform the two-particle FD projective
measurements on their own particles $(1,4)$ and $(2,5)$,
respectively. The measurement basis chosen by Alice$_{1}$ and
Alice$_{2}$ are the set of MOBVs $\{|\psi_{gh}^{(k)}\rangle\}$
$(g,h=0,1,2,3, k=1,2)$
\begin{eqnarray}\label{eq8}
|\psi_{00}^{(k)}\rangle=a_{k}|00\rangle+b_{k}|11\rangle+c_{k}|22\rangle+d_{k}|33\rangle,\ \ \ \nonumber\\
|\psi_{10}^{(k)}\rangle=b_{k}|00\rangle-a_{k}|11\rangle+d_{k}|22\rangle-c_{k}|33\rangle,\ \ \ \nonumber\\
|\psi_{20}^{(k)}\rangle=-c_{k}|00\rangle+d_{k}|11\rangle+a_{k}|22\rangle-b_{k}|33\rangle,\nonumber\\
|\psi_{30}^{(k)}\rangle=-d_{k}|00\rangle-c_{k}|11\rangle+b_{k}|22\rangle+a_{k}|33\rangle,\nonumber\\
|\psi_{01}^{(k)}\rangle=a_{k}|01\rangle+b_{k}|12\rangle+c_{k}|23\rangle+d_{k}|30\rangle,\ \ \ \nonumber\\
|\psi_{11}^{(k)}\rangle=b_{k}|01\rangle-a_{k}|12\rangle+d_{k}|23\rangle-c_{k}|30\rangle,\ \ \ \nonumber\\
|\psi_{21}^{(k)}\rangle=-c_{k}|01\rangle+d_{k}|12\rangle+a_{k}|23\rangle-b_{k}|30\rangle,\nonumber\\
|\psi_{31}^{(k)}\rangle=-d_{k}|01\rangle-c_{k}|12\rangle+b_{k}|23\rangle+a_{k}|30\rangle,\nonumber\\
|\psi_{02}^{(k)}\rangle=a_{k}|02\rangle+b_{k}|13\rangle+c_{k}|20\rangle+d_{k}|31\rangle,\ \ \ \nonumber\\
|\psi_{12}^{(k)}\rangle=b_{k}|02\rangle-a_{k}|13\rangle+d_{k}|20\rangle-c_{k}|31\rangle,\ \ \ \nonumber\\
|\psi_{22}^{(k)}\rangle=-c_{k}|02\rangle+d_{k}|13\rangle+a_{k}|20\rangle-b_{k}|31\rangle,\nonumber\\
|\psi_{32}^{(k)}\rangle=-d_{k}|02\rangle-c_{k}|13\rangle+b_{k}|20\rangle+a_{k}|31\rangle,\nonumber\\
|\psi_{03}^{(k)}\rangle=a_{k}|03\rangle+b_{k}|10\rangle+c_{k}|21\rangle+d_{k}|32\rangle,\ \ \ \nonumber\\
|\psi_{13}^{(k)}\rangle=b_{k}|03\rangle-a_{k}|10\rangle+d_{k}|21\rangle-c_{k}|32\rangle,\ \ \ \nonumber\\
|\psi_{23}^{(k)}\rangle=-c_{k}|03\rangle+d_{k}|10\rangle+a_{k}|21\rangle-b_{k}|32\rangle,\nonumber\\
|\psi_{33}^{(k)}\rangle=-d_{k}|03\rangle-c_{k}|10\rangle+b_{k}|21\rangle+a_{k}|32\rangle,
\end{eqnarray}
where $\{|\psi_{gh}^{(1)}\rangle\}$ is a set of MOBVs chosen by
Alice$_{1}$, and $\{|\psi_{gh}^{(2)}\rangle\}$ by Alice$_{2}$. From
Eq.(8), the quantum channel composed of entangled states (7) can be
written as
\begin{eqnarray}\label{eq9}
|\Phi\rangle=|\phi\rangle_{123}\otimes|\phi\rangle_{456}\qquad\qquad\qquad\qquad\qquad\qquad\qquad\nonumber\\
=\frac{1}{4}[|\psi_{00}^{(1)}\rangle_{14}|\psi_{00}^{(2)}\rangle_{25}(a|00\rangle+b|11\rangle+c|22\rangle+d|33\rangle)_{36}\nonumber\\
+|\psi_{10}^{(1)}\rangle_{14}|\psi_{10}^{(2)}\rangle_{25}(b|00\rangle+a|11\rangle+d|22\rangle+c|33\rangle)_{36}\nonumber\\
+|\psi_{20}^{(1)}\rangle_{14}|\psi_{20}^{(2)}\rangle_{25}(c|00\rangle+d|11\rangle+a|22\rangle+b|33\rangle)_{36}\nonumber\\
+|\psi_{30}^{(1)}\rangle_{14}|\psi_{30}^{(2)}\rangle_{25}(d|00\rangle+c|11\rangle+b|22\rangle+a|33\rangle)_{36}\nonumber\\
+|\psi_{01}^{(1)}\rangle_{14}|\psi_{01}^{(2)}\rangle_{25}(a|01\rangle+b|12\rangle+c|23\rangle+d|30\rangle)_{36}\nonumber\\
+|\psi_{11}^{(1)}\rangle_{14}|\psi_{11}^{(2)}\rangle_{25}(b|01\rangle+a|12\rangle+d|23\rangle+c|30\rangle)_{36}\nonumber\\
+|\psi_{21}^{(1)}\rangle_{14}|\psi_{21}^{(2)}\rangle_{25}(c|01\rangle+d|12\rangle+a|23\rangle+b|30\rangle)_{36}\nonumber\\
+|\psi_{31}^{(1)}\rangle_{14}|\psi_{31}^{(2)}\rangle_{25}(d|01\rangle+c|12\rangle+b|23\rangle+a|30\rangle)_{36}\nonumber\\
+|\psi_{02}^{(1)}\rangle_{14}|\psi_{02}^{(2)}\rangle_{25}(a|02\rangle+b|13\rangle+c|20\rangle+d|31\rangle)_{36}\nonumber\\
+|\psi_{12}^{(1)}\rangle_{14}|\psi_{12}^{(2)}\rangle_{25}(b|02\rangle+a|13\rangle+d|20\rangle+c|31\rangle)_{36}\nonumber\\
+|\psi_{22}^{(1)}\rangle_{14}|\psi_{22}^{(2)}\rangle_{25}(c|02\rangle+d|13\rangle+a|20\rangle+b|31\rangle)_{36}\nonumber\\
+|\psi_{32}^{(1)}\rangle_{14}|\psi_{32}^{(2)}\rangle_{25}(d|02\rangle+c|13\rangle+b|20\rangle+a|31\rangle)_{36}\nonumber\\
+|\psi_{03}^{(1)}\rangle_{14}|\psi_{03}^{(2)}\rangle_{25}(a|03\rangle+b|10\rangle+c|21\rangle+d|32\rangle)_{36}\nonumber\\
+|\psi_{13}^{(1)}\rangle_{14}|\psi_{13}^{(2)}\rangle_{25}(b|03\rangle+a|10\rangle+d|21\rangle+c|32\rangle)_{36}\nonumber\\
+|\psi_{23}^{(1)}\rangle_{14}|\psi_{23}^{(2)}\rangle_{25}(c|03\rangle+d|10\rangle+a|21\rangle+b|32\rangle)_{36}\nonumber\\
+|\psi_{33}^{(1)}\rangle_{14}|\psi_{33}^{(2)}\rangle_{25}(d|03\rangle+c|10\rangle+b|21\rangle+a|32\rangle)_{36}\nonumber\\
+|\psi_{00}^{(1)}\rangle_{14}|\psi_{10}^{(2)}\rangle_{25}(a_{1}b_{2}|00\rangle-a_{2}b_{1}|11\rangle+c_{1}d_{2}|22\rangle-c_{2}d_{1}|33\rangle)_{36}+\cdots
\ ],
\end{eqnarray}
where "$\cdots$" includes 47 other terms with $g\neq m$ or/and
$h\neq n$ in
$|\psi_{gh}^{(1)}\rangle_{14}|\psi_{mn}^{(2)}\rangle_{25}$
$(g,h,m,n=0,1,2,3)$. In Eq.(8) only the 16 first terms give rise to
a success, all the 48 remaining terms with $g\neq m$ or/and $h\neq
n$ lead to a failure. Now let Alice$_{1}$ and Alice$_{2}$ perform
the two-particle FD projective measurements on their own particles
$(1,4)$ and $(2,5)$, respectively, and then they inform Bob of their
outcomes in public. In accord with the measurement outcomes of
Alice$_{1}$ and Alice$_{2}$, Bob can reconstruct the original state.
For instance, suppose Alice$_{1}$'s measurement outcome is
$|\psi_{11}^{(1)}\rangle_{14}$ and Alice$_{2}$'s outcome is
$|\psi_{11}^{2}\rangle_{25}$, the particles 3 and 6 will collapse
into the state
$\frac{1}{4}(b|01\rangle+a|12\rangle+d|23\rangle+c|30\rangle)_{36}$.
According to Alice$_{1}$'s and Alice$_{2}$'s public announcement,
Bob should perform the unitary operations $U_{1}\otimes U_{5}$ on
particles 3 and 6, thus the bipartite FD entangled state (6) can be
reconstructed. Here unitary operation $U_{1}$ is defined by Eq.(5)
and $U_{5}$ is one of the unitary operations $U_{j}$ $(j=4,5,6,7)$
\begin{eqnarray}\label{eq10}
U_{4}=\left(
\begin{array}{cccc}
 0 & 1 & 0 & 0 \\
 0 & 0 & 1 & 0\\
 0 & 0 & 0 & 1\\
 1 & 0 & 0 & 0
\end{array}
\right),\quad U_{5}=\left(
\begin{array}{cccc}
0 & 0 & 1 & 0 \\
0 & 1 & 0 & 0 \\
1 & 0 & 0 & 0\\
0 & 0 & 0 & 1
\end{array}
\right),\nonumber\\  U_{6}=\left(
\begin{array}{cccc}
0 & 0 & 0 & 1\\
1 & 0 & 0 & 0\\
0 & 1 & 0 & 0 \\
0 & 0 & 1 & 0
\end{array}
\right),\quad U_{7}=\left(
\begin{array}{cccc}
 1 & 0 & 0 & 0\\
0 & 0 & 0 & 1\\
0 & 0 & 1 & 0\\
0 & 1 & 0 & 0
\end{array}
\right).
\end{eqnarray}
\begin{table}
\caption{Corresponding relation between the measurement results (MR)
of Alice$_{1}$ and Alice$_{2}$ and the local unitary operations
$(U_{i})_{3}\otimes (U_{j})_{6}$ $(i,j=0,1,\cdots,7)$ performed by
Bob. ($\zeta_{gh}\rightarrow |\psi_{gh}^{(1)}\rangle_{14}$,
$\eta_{mn}\rightarrow |\psi_{mn}^{(2)}\rangle_{25}$,
$u_{i}\rightarrow (U_{i})_{3}$, $v_{j}\rightarrow(U_{j})_{6}$,
$g,h,m,n=0,1,2,3$) }\label{tab2}
\begin{center}
\begin{tabular}{cccc} \hline
\quad MR & \qquad $u_{i}\otimes v_{j}$ &\qquad MR & \qquad $u_{i}\otimes v_{j}$\\
\hline
\quad$\zeta_{00}\eta_{00}$ &\qquad $u_{0}\otimes v_{0}$ &\qquad $\zeta_{02}\eta_{02}$ &\qquad $u_{0}\otimes v_{2}$\\
\quad$\zeta_{10}\eta_{10}$ &\qquad $u_{1}\otimes v_{1}$ &\qquad $\zeta_{12}\eta_{12}$ &\qquad $u_{1}\otimes v_{3}$\\
\quad$\zeta_{20}\eta_{20}$ &\qquad $u_{2}\otimes v_{2}$ &\qquad $\zeta_{22}\eta_{22}$ &\qquad $u_{2}\otimes v_{0}$\\
\quad$\zeta_{30}\eta_{30}$ &\qquad $u_{3}\otimes v_{3}$ &\qquad $\zeta_{32}\eta_{32}$ &\qquad $u_{3}\otimes v_{1}$\\
\quad$\zeta_{01}\eta_{01}$ &\qquad $u_{1}\otimes v_{4}$ &\qquad $\zeta_{03}\eta_{03}$ &\qquad $u_{0}\otimes v_{6}$\\
\quad$\zeta_{11}\eta_{11}$ &\qquad $u_{1}\otimes v_{5}$ &\qquad $\zeta_{13}\eta_{13}$ &\qquad $u_{1}\otimes v_{7}$\\
\quad$\zeta_{21}\eta_{21}$ &\qquad $u_{2}\otimes v_{6}$ &\qquad $\zeta_{23}\eta_{23}$ &\qquad $u_{2}\otimes v_{4}$\\
\quad$\zeta_{31}\eta_{31}$ &\qquad $u_{3}\otimes v_{7}$ &\qquad $\zeta_{33}\eta_{33}$ &\qquad $u_{3}\otimes v_{5}$\\
\hline
\end{tabular}
\end{center}
\end{table}
If the measurement outcomes of Alice$_{1}$ and Alice$_{2}$ are the
other 15 cases of the sixteen first terms in Eq.(9), the relation
between the outcomes by Alice$_{1}$ and Alice$_{2}$ and the unitary
operations by Bob is shown in Table 2. The required classical
communication cost is 8 bits in this protocol.

\subsection{JRSP by using three bipartite FD entangled states as the quantum channel}
Suppose the state that Alice$_{1}$ and Alice$_{2}$ wish to help Bob
remotely prepare is still in state $|\psi\rangle$ (see Eq.(6)). We
also assume that Alice$_{1}$, Alice$_{2}$ and Bob share three
bipartite FD entangled states as quantum channel
\begin{eqnarray}\label{eq11}
|\phi\rangle_{12}=\frac{1}{2}\sum_{j=0}^{3}|jj\rangle_{12}\nonumber\\
|\phi\rangle_{34}=\frac{1}{2}\sum_{j=0}^{3}|jj\rangle_{34}\nonumber\\
|\phi\rangle_{56}=\frac{1}{2}\sum_{j=0}^{3}|jj\rangle_{56},
\end{eqnarray}
where particles 1 and 3 belong to Alice$_{1}$, particles 2 and 5 to
Alice$_{2}$ and particles 4 and 6 to Bob, respectively. As in the
previous protocol, Alice$_{1}$ and Alice$_{2}$ perform the
two-particle FD projective measurements on their own particles
$(1,3)$ and $(2,5)$, respectively. The measurement basis chosen by
Alice$_{1}$ and Alice$_{2}$ is still in $\{|\psi_{gh}^{k}\rangle\}$
(see Eq.(8)). The quantum channel
$|\Phi\rangle=|\phi\rangle_{12}|\phi\rangle_{34}|\phi\rangle_{56}$
can be written in terms of basis $\{|\psi_{gh}^{k}\rangle\}$ as
\begin{eqnarray}\label{eq12}
|\Psi\rangle=\frac{1}{8}\sum_{j=0}^{3}[|G_{0j}\rangle(a|\lambda_{0j}\rangle+b|\lambda_{1j}\rangle+c|\lambda_{2j}\rangle+d|\lambda_{3j}\rangle)_{46}\nonumber\\
+|G_{1j}\rangle(b|\lambda_{0j}\rangle+a|\lambda_{1j}\rangle+d|\lambda_{2j}\rangle+c|\lambda_{3j}\rangle)_{46}\nonumber\\
+|G_{2j}\rangle(c|\lambda_{0j}\rangle+d|\lambda_{1j}\rangle+a|\lambda_{2j}\rangle+b|\lambda_{3j}\rangle)_{46}\nonumber\\
+|G_{3j}\rangle(d|\lambda_{0j}\rangle+c|\lambda_{2j}\rangle+b|\lambda_{2j}\rangle+a|\lambda_{3j}\rangle)_{46}\nonumber\\
+|G_{4j}\rangle(a|\lambda_{1j}\rangle+b|\lambda_{2j}\rangle+c|\lambda_{3j}\rangle+d|\lambda_{0j}\rangle)_{46}\nonumber\\
+|G_{5j}\rangle(b|\lambda_{1j}\rangle+a|\lambda_{2j}\rangle+d|\lambda_{3j}\rangle+c|\lambda_{0j}\rangle)_{46}\nonumber\\
+|G_{6j}\rangle(c|\lambda_{1j}\rangle+d|\lambda_{2j}\rangle+a|\lambda_{3j}\rangle+b|\lambda_{0j}\rangle)_{46}\nonumber\\
+|G_{7j}\rangle(d|\lambda_{1j}\rangle+c|\lambda_{2j}\rangle+b|\lambda_{3j}\rangle+a|\lambda_{0j}\rangle)_{46}\nonumber\\
+|\psi_{00}^{(1)}\rangle_{13}|\psi_{10}^{(2)}\rangle_{25}(a_{1}b_{2}|00\rangle-b_{1}a_{2}|11\rangle+c_{1}d_{2}|22\rangle-d_{1}c_{2}|33\rangle)_{46}\nonumber\\
+\cdots \ ],
\end{eqnarray}
where $|G_{pj}\rangle\equiv|G_{pj}\rangle_{1325}$ given in appendix
A and $|G_{0j}\rangle\sim |G_{7j}\rangle$ include 64 terms with
$g=m$ or/and $h\neq n$ in
$|\psi_{gh}^{(1)}\rangle_{13}|\psi_{mn}^{(2)}\rangle_{25}$
$(g,h,m,n=0,1,2,3)$, "$\cdots$" includes 191 other terms with $g\neq
m$ or/and $h\neq n$ in
$|\psi_{gh}^{(1)}\rangle_{13}|\psi_{mn}^{(2)}\rangle_{25}$,
$|\lambda_{ij}\rangle\equiv |i,i\oplus j\rangle$ and $i\oplus j$
means $i+j$ mod 3. In Eq.(12) only the 64 first terms (i.e.
$|G_{0j\rangle}\sim|G_{7j}\rangle$) can cause success, but all the
192 remaining terms with $g\neq m$ lead to failure. For example,
assume Alice$_{1}$'s measurement result is
$|\psi_{20}^{(1)}\rangle_{13}$ and Alice$_{2}$'s result is
$|\psi_{21}^{2}\rangle_{25}$, the particles 4 and 6 will collapse
into the state
$\frac{1}{8}(c|01\rangle+d|12\rangle+a|23\rangle+b|30\rangle)_{46}$,
then Bob should perform $U_{2}\otimes U_{6}$ on particles 4 and 6,
the original state(6) can be reconstructed successfully. If the
measurement results of Alice$_{1}$ and Alice$_{2}$ are the other 63
cases of the successful terms in Eq.(12), the relation between the
results by Alice$_{1}$ and Alice$_{2}$ and the unitary operations by
Bob is shown in Table 3. In this protocol, the required classical
communication cost is also 8 bits.

\begin{table}
\caption{ Corresponding relation between the measurement results
(MR) of Alice$_{1}$ and Alice$_{2}$ and the local unitary operations
$(U_{i})_{4}\otimes (U_{j})_{6}$ $(i,j=0,1,\cdots,7)$ by Bob.
($\xi_{gh}\rightarrow |\psi_{gh}^{(1)}\rangle_{13}$,
$\tau_{mn}\rightarrow |\psi_{mn}^{(2)}\rangle_{25}$,
$u_{i}\rightarrow (U_{i})_{4}$, $v_{j}\rightarrow(U_{j})_{6}$,
$g,h,m,n=0,1,2,3$). }\label{tab3}
\begin{center}
\begin{tabular*}{140mm}{cccc} \hline
\quad MR & \qquad $u_{i}\otimes v_{j}$ &\qquad MR & \qquad $u_{i}\otimes v_{j}$\\
\hline
\quad$\xi_{00}\tau_{00}(or\ \xi_{22}\tau_{22})$ &\qquad $u_{0}\otimes v_{0}$ &\qquad $\xi_{01}\tau_{00}(or\ \xi_{23}\tau_{22})$ &\qquad $u_{4}\otimes v_{0}$\\
\quad$\xi_{00}\tau_{01}(or\ \xi_{22}\tau_{23})$ &\qquad $u_{0}\otimes v_{4}$ &\qquad $\xi_{01}\tau_{01}(or\ \xi_{23}\tau_{23})$ &\qquad $u_{4}\otimes v_{4}$\\
\quad$\xi_{00}\tau_{02}(or\ \xi_{22}\tau_{20})$ &\qquad $u_{0}\otimes v_{2}$ &\qquad $\xi_{01}\tau_{02}(or\ \xi_{23}\tau_{20})$ &\qquad $u_{4}\otimes v_{2}$\\
\quad$\xi_{00}\tau_{03}(or\ \xi_{22}\tau_{21})$ &\qquad $u_{0}\otimes v_{6}$ &\qquad $\xi_{01}\tau_{03}(or\ \xi_{23}\tau_{21})$ &\qquad $u_{4}\otimes v_{6}$\\
\quad$\xi_{10}\tau_{10}(or\ \xi_{32}\tau_{32})$ &\qquad $u_{1}\otimes v_{1}$ &\qquad $\xi_{11}\tau_{10}(or\ \xi_{33}\tau_{32})$ &\qquad $u_{5}\otimes v_{1}$\\
\quad$\xi_{10}\tau_{11}(or\ \xi_{32}\tau_{33})$ &\qquad $u_{1}\otimes v_{5}$ &\qquad $\xi_{11}\tau_{11}(or\ \xi_{33}\tau_{33})$ &\qquad $u_{5}\otimes v_{5}$\\
\quad$\xi_{10}\tau_{12}(or\ \xi_{32}\tau_{30})$ &\qquad $u_{1}\otimes v_{3}$ &\qquad $\xi_{11}\tau_{12}(or\ \xi_{33}\tau_{30})$ &\qquad $u_{5}\otimes v_{3}$\\
\quad$\xi_{10}\tau_{13}(or\ \xi_{32}\tau_{31})$ &\qquad $u_{1}\otimes v_{7}$ &\qquad $\xi_{11}\tau_{13}(or\ \xi_{33}\tau_{31})$ &\qquad $u_{5}\otimes v_{7}$\\
\quad$\xi_{20}\tau_{20}(or\ \xi_{02}\tau_{02})$ &\qquad $u_{2}\otimes v_{2}$ &\qquad $\xi_{21}\tau_{20}(or\ \xi_{03}\tau_{02})$ &\qquad $u_{6}\otimes v_{2}$\\
\quad$\xi_{20}\tau_{21}(or\ \xi_{02}\tau_{03})$ &\qquad $u_{2}\otimes v_{6}$ &\qquad $\xi_{21}\tau_{21}(or\ \xi_{03}\tau_{03})$ &\qquad $u_{6}\otimes v_{6}$\\
\quad$\xi_{20}\tau_{22}(or\ \xi_{02}\tau_{00})$ &\qquad $u_{2}\otimes v_{0}$ &\qquad $\xi_{21}\tau_{22}(or\ \xi_{03}\tau_{00})$ &\qquad $u_{6}\otimes v_{0}$\\
\quad$\xi_{20}\tau_{23}(or\ \xi_{02}\tau_{01})$ &\qquad $u_{2}\otimes v_{4}$ &\qquad $\xi_{21}\tau_{23}(or\ \xi_{03}\tau_{01})$ &\qquad $u_{6}\otimes v_{4}$\\
\quad$\xi_{30}\tau_{30}(or\ \xi_{12}\tau_{12})$ &\qquad $u_{3}\otimes v_{3}$ &\qquad $\xi_{31}\tau_{30}(or\ \xi_{13}\tau_{12})$ &\qquad $u_{7}\otimes v_{3}$\\
\quad$\xi_{30}\tau_{31}(or\ \xi_{12}\tau_{13})$ &\qquad $u_{3}\otimes v_{7}$ &\qquad $\xi_{31}\tau_{31}(or\ \xi_{13}\tau_{13})$ &\qquad $u_{7}\otimes v_{7}$\\
\quad$\xi_{30}\tau_{32}(or\ \xi_{12}\tau_{10})$ &\qquad $u_{3}\otimes v_{1}$ &\qquad $\xi_{31}\tau_{32}(or\ \xi_{13}\tau_{10})$ &\qquad $u_{7}\otimes v_{1}$\\
\quad$\xi_{30}\tau_{33}(or\ \xi_{12}\tau_{11})$ &\qquad $u_{3}\otimes v_{5}$ &\qquad $\xi_{31}\tau_{33}(or\ \xi_{13}\tau_{11})$ &\qquad $u_{7}\otimes v_{5}$\\
\hline
\end{tabular*}
\end{center}
\end{table}

\section{Conclusion}
We propose the protocols for joint remote preparation of the
four-dimensional quantum states by using various types of the
four-dimensional entangled states as quantum channel. In these
protocols, two senders share an original state which they wish to
help the receiver to remotely prepare it, but each sender only
partly knows the state. It is shown that, only if when all the
senders collaborate with each other, the receiver can remotely
reconstruct the original state. In order to realize the JRSP, two
senders need to perform four-dimensional projective measurements on
their own particle, respectively, and then inform the receiver Bob
of the measurement outcomes through the classical channel. According
to the public information of the senders, the receiver can obtain
the original state by using some appropriate unitary operations.
These protocols require resources such as bipartite or tripartite
four-dimensional entangled state as the quantum channel, single- or
two-particle four-dimensional projective measurement, classical
communication and appropriate unitary operation. In principle, our
protocols can be generalized to the case of JRSP of
\emph{d}-dimensional ($d=2^{N}$, $N$ is a positive integer greater
than 2) quantum state. Furthermore, the required classical
communication cost in the JRSP process in our protocols has been
calculated respectively.

\newpage
\appendix{\textbf{Appendix A.}}
\bigskip

The states $|G_{pj}\rangle$ $(p=0,1,\cdots,7; j=0,1,2,3)$ in Eq.(12)
are of the form
\renewcommand{\theequation}{A. \arabic{equation}}
\begin{eqnarray}\label{A.1}
|G_{00}\rangle=|\psi_{00}^{1}\rangle_{13}|\psi_{00}^{2}\rangle_{25}+|\psi_{22}^{1}\rangle_{13}|\psi_{22}^{2}\rangle_{25},\\
|G_{01}\rangle=|\psi_{00}^{1}\rangle_{13}|\psi_{01}^{2}\rangle_{25}+|\psi_{22}^{1}\rangle_{13}|\psi_{23}^{2}\rangle_{25},\\
|G_{02}\rangle=|\psi_{00}^{1}\rangle_{13}|\psi_{02}^{2}\rangle_{25}+|\psi_{22}^{1}\rangle_{13}|\psi_{20}^{2}\rangle_{25},\\
|G_{03}\rangle=|\psi_{00}^{1}\rangle_{13}|\psi_{03}^{2}\rangle_{25}+|\psi_{22}^{1}\rangle_{13}|\psi_{21}^{2}\rangle_{25},\\
|G_{10}\rangle=|\psi_{10}^{1}\rangle_{13}|\psi_{10}^{2}\rangle_{25}+|\psi_{32}^{1}\rangle_{13}|\psi_{32}^{2}\rangle_{25},\\
|G_{11}\rangle=|\psi_{10}^{1}\rangle_{13}|\psi_{11}^{2}\rangle_{25}+|\psi_{32}^{1}\rangle_{13}|\psi_{33}^{2}\rangle_{25},\\
|G_{12}\rangle=|\psi_{10}^{1}\rangle_{13}|\psi_{12}^{2}\rangle_{25}+|\psi_{32}^{1}\rangle_{13}|\psi_{30}^{2}\rangle_{25},\\
|G_{13}\rangle=|\psi_{10}^{1}\rangle_{13}|\psi_{13}^{2}\rangle_{25}+|\psi_{32}^{1}\rangle_{13}|\psi_{31}^{2}\rangle_{25},\\
|G_{20}\rangle=|\psi_{20}^{1}\rangle_{13}|\psi_{20}^{2}\rangle_{25}+|\psi_{02}^{1}\rangle_{13}|\psi_{02}^{2}\rangle_{25},\\
|G_{21}\rangle=|\psi_{20}^{1}\rangle_{13}|\psi_{21}^{2}\rangle_{25}+|\psi_{02}^{1}\rangle_{13}|\psi_{03}^{2}\rangle_{25},\\
|G_{22}\rangle=|\psi_{20}^{1}\rangle_{13}|\psi_{22}^{2}\rangle_{25}+|\psi_{02}^{1}\rangle_{13}|\psi_{00}^{2}\rangle_{25},\\
|G_{23}\rangle=|\psi_{20}^{1}\rangle_{13}|\psi_{23}^{2}\rangle_{25}+|\psi_{02}^{1}\rangle_{13}|\psi_{01}^{2}\rangle_{25},\\
|G_{30}\rangle=|\psi_{30}^{1}\rangle_{13}|\psi_{30}^{2}\rangle_{25}+|\psi_{12}^{1}\rangle_{13}|\psi_{12}^{2}\rangle_{25},\\
|G_{31}\rangle=|\psi_{30}^{1}\rangle_{13}|\psi_{31}^{2}\rangle_{25}+|\psi_{12}^{1}\rangle_{13}|\psi_{13}^{2}\rangle_{25},\\
|G_{32}\rangle=|\psi_{30}^{1}\rangle_{13}|\psi_{32}^{2}\rangle_{25}+|\psi_{12}^{1}\rangle_{13}|\psi_{10}^{2}\rangle_{25},\\
|G_{33}\rangle=|\psi_{30}^{1}\rangle_{13}|\psi_{33}^{2}\rangle_{25}+|\psi_{12}^{1}\rangle_{13}|\psi_{11}^{2}\rangle_{25},\\
|G_{40}\rangle=|\psi_{01}^{1}\rangle_{13}|\psi_{00}^{2}\rangle_{25}+|\psi_{23}^{1}\rangle_{13}|\psi_{22}^{2}\rangle_{25},\\
|G_{41}\rangle=|\psi_{01}^{1}\rangle_{13}|\psi_{01}^{2}\rangle_{25}+|\psi_{23}^{1}\rangle_{13}|\psi_{23}^{2}\rangle_{25},\\
|G_{42}\rangle=|\psi_{01}^{1}\rangle_{13}|\psi_{02}^{2}\rangle_{25}+|\psi_{23}^{1}\rangle_{13}|\psi_{20}^{2}\rangle_{25},\\
|G_{43}\rangle=|\psi_{01}^{1}\rangle_{13}|\psi_{03}^{2}\rangle_{25}+|\psi_{23}^{1}\rangle_{13}|\psi_{21}^{2}\rangle_{25},\\
|G_{50}\rangle=|\psi_{11}^{1}\rangle_{13}|\psi_{10}^{2}\rangle_{25}+|\psi_{33}^{1}\rangle_{13}|\psi_{32}^{2}\rangle_{25},\\
|G_{51}\rangle=|\psi_{11}^{1}\rangle_{13}|\psi_{11}^{2}\rangle_{25}+|\psi_{33}^{1}\rangle_{13}|\psi_{33}^{2}\rangle_{25},\\
|G_{52}\rangle=|\psi_{11}^{1}\rangle_{13}|\psi_{12}^{2}\rangle_{25}+|\psi_{33}^{1}\rangle_{13}|\psi_{30}^{2}\rangle_{25},\\
|G_{53}\rangle=|\psi_{11}^{1}\rangle_{13}|\psi_{13}^{2}\rangle_{25}+|\psi_{33}^{1}\rangle_{13}|\psi_{31}^{2}\rangle_{25},\\
|G_{60}\rangle=|\psi_{21}^{1}\rangle_{13}|\psi_{20}^{2}\rangle_{25}+|\psi_{03}^{1}\rangle_{13}|\psi_{02}^{2}\rangle_{25},\\
|G_{61}\rangle=|\psi_{21}^{1}\rangle_{13}|\psi_{21}^{2}\rangle_{25}+|\psi_{03}^{1}\rangle_{13}|\psi_{03}^{2}\rangle_{25},\\
|G_{62}\rangle=|\psi_{21}^{1}\rangle_{13}|\psi_{22}^{2}\rangle_{25}+|\psi_{03}^{1}\rangle_{13}|\psi_{00}^{2}\rangle_{25},\\
|G_{63}\rangle=|\psi_{21}^{1}\rangle_{13}|\psi_{23}^{2}\rangle_{25}+|\psi_{03}^{1}\rangle_{13}|\psi_{01}^{2}\rangle_{25},\\
|G_{70}\rangle=|\psi_{31}^{1}\rangle_{13}|\psi_{30}^{2}\rangle_{25}+|\psi_{13}^{1}\rangle_{13}|\psi_{12}^{2}\rangle_{25},\\
|G_{71}\rangle=|\psi_{31}^{1}\rangle_{13}|\psi_{31}^{2}\rangle_{25}+|\psi_{13}^{1}\rangle_{13}|\psi_{13}^{2}\rangle_{25},\\
|G_{72}\rangle=|\psi_{31}^{1}\rangle_{13}|\psi_{32}^{2}\rangle_{25}+|\psi_{13}^{1}\rangle_{13}|\psi_{10}^{2}\rangle_{25},\\
|G_{73}\rangle=|\psi_{31}^{1}\rangle_{13}|\psi_{33}^{2}\rangle_{25}+|\psi_{13}^{1}\rangle_{13}|\psi_{11}^{2}\rangle_{25}.
\end {eqnarray}

\bigskip\bigskip

\end{document}